# Can a closed critical surface in a quark-gluon plasma serve as a model for the behavior of quantum gravity near to an event horizon?

George Chapline
Lawrence Livermore National Laboratory

Time stands still at a quantum critical point in the sense that correlation functions near to the critical point are approximately independent of frequency. In the case of a quantum liquid this would imply that classical hydrodynamics breaks down near to the critical point, revealing the underlying quantum degrees of freedom. An opportunity to see this effect for the first time in the laboratory may be provided by relativistic heavy ion collisions that are tuned so that the quark-gluon plasma passes through its critical point forming a closed critical surface. In this note we point out that in certain kinds of quantum fluids the temperature of a spherical critical surface will be proportional to (radius)$^{-1}$ and the entropy inside the surface will be close to the Bekenstein bound. In these cases the breakdown in hydrodynamics near to the critical point might serve as a model for the behavior of quantum gravity near to an event horizon. Such a possibility is *a fortiori* notable because general relativity predicts that nothing should happen at an event horizon.

Although it was originally believed that high temperature quark-gluon plasma (QGP) consists of an almost ideal gas of quarks and gluons [1], it is now clear that this simple picture is not correct. In particular, one drammatic departure from what would be expected in an idea gas is the very likely existence of a quantum critical point in the phase diagram of hadronic matter. Although the quark-gluon plasmas that have to date been created via relativistic heavy ion collisions do not appear to pass through a critical point, the existence of a quantum critical point can be inferred by comparing the experimental data on relativistic heavy ion collisions with lattice gauge calculations and a hadron resonance model for hadronic matter [2]. In this paper we wish first of all to point out that relativistic heavy ion collisions where the initial conditions are chosen so that the path of the QGP passes through the critical point are interesting because of what may be revealed regarding the fundamental degrees of freedom in a hot QGP. Our main aim is to draw attention to the possibility that the appearance of quantum degrees of freedom close to a QGP critical surface might serve as a model for the emergence of quantum effects near to where classical general relativity would predict that an event horizon would form.

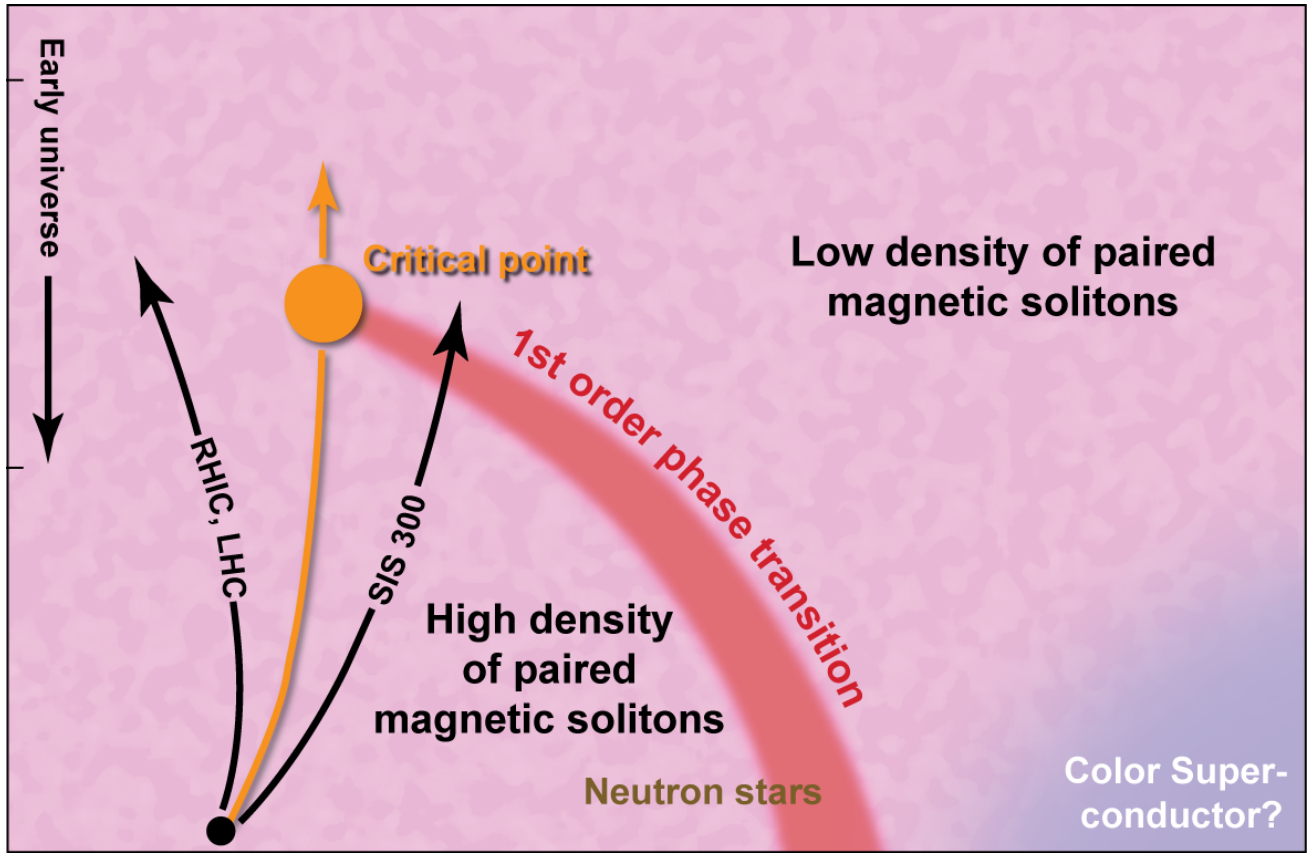

Fig. 1. Schematic phase diagram for quark matter where what are usually identified as the hadronic and quark-gluon plasma phases are combined into one phase with varying densities of paired magnetic solitons. Also shown is a typical path in equation of state space for the quark-gluon plasmas that have been created in the recent past using the RHIC and LHC accelerators. Future experiments using the SIS 300 accelerator may be able to access the phase transition line where the density of paired magnetic solitons changes rapidly, and perhaps even the critical point.

Although a QGP is not a superfluid, analyses of RHIC collisions [3] as well as AdS/CFT duality [4] suggest that the viscosity of a high temperature QGP is relatively small. Therefore it may be a reasonable approximation to ignore classical dissipation, and use a nonlinear Schrodinger equation to describe the low frequency dynamics of the QGP. We will be particularly interested in the dynamics of a hot QGP where the initial conditions for its creation have been specially chosen so as to allow the path of the QGP to pass through its critical point (Fig. 1). One may then imagine that at the time of maximum compression the quark-gluon plasma will contain a closed critical suface. If we regard the QGP as consisting of a fluid of Bose particles with mass $m$ (we discuss below the validity of this assumption), then near to the critical surface the dynamics of the QGP can be described by an effective Lagrangian of the form [5]

$$L_{eff} = \psi^*\left(i\hbar\frac{\partial}{\partial t}+\mu\right)\psi - \frac{\hbar^2}{2m}|\nabla\psi|^2 - \frac{1}{2}mc_s^2\left(|\psi|-|\psi_0|\right)^2 - \frac{3P_c}{\rho_c^2}\left(|\psi|-|\psi_0|\right)^4, \tag{1}$$

where $c_s$ is the speed of sound, and $\rho_c$ and $P_c$ are the density and pressure at the critical point. It follows from (1) that mean field small amplitude perturbations $\phi \equiv \delta\langle\mathrm{Re}\psi\rangle$ approaching the critical surface will be described by an acoustic equation of the form :

$$\hbar^2\frac{\partial^2\phi}{\partial t^2} = \frac{\partial^2\phi}{\partial^2 t} = \frac{1}{\tau_0^2}\nabla^2(z^2\phi) - \left(\frac{\hbar}{2m}\right)^2\nabla^4\phi . \tag{2}$$

where we have assumed that the speed of sound $c_s$ goes to zero linearly as the critical surface is approached; i.e. $c_s = z/\tau_0$. Qualitatively one expects that because small amplitude perturbations no longer propagate ballistically as $z \to 0$, the evolution of a disturbance on the critical surface will be largely determined by the second term on the r.h.s. of Eq. (1), leading to a spreading of the wave function over the surface of the critical surface.

Eq. (2) implies that small amplitude periodic waves approaching the critical surface have a dispersion relation [5]:

$$\hbar\omega_q = \sqrt{(\hbar q c_s)^2 + \left(\frac{\hbar^2 q^2}{2m}\right)^2} , \tag{3}$$

Therefore within the framework of the Bose fluid model small amplitude collective excitations will morph into heavy non-relativistic particles with mass $m$ when they are within a distance

$$z^* \approx R_c \sqrt{\hbar\omega/2mc_s^2}, \qquad (4)$$

from the critical surface, where $R_c$ is the radius of the critical surface. This behavior is typical feature of quantum critical phase transitions in that collective macroscopic collective degrees of fredom will in general degrade into particle-like microscopic degrees of freedom near to a quantum critical point [6]. As a consequence, a classical hydrodynamic description of the flow will fail for length scales smaller than $\hbar/mc_s$ and frequencies $\omega > mc_s^2/\hbar$. As $z \to 0$ hydrodynamics is a good approximation only in the limit $\omega \to 0$. Therefore, whereas classical hydrodynamics may have worked well for describing the dynamics of QGPs that to date have been created using the RHIC and LHC (see e.g. [3]), we expect that classical hydrodynamics will be inadequate for describing the dynamics of a QGP in which a closed critical surface forms.

Actually the mean field approximation itself breaks down close to a critical surface due to large quantum fluctuations. If we second quantize the effective Lagrangian (1), then one can use Feynman diagrams to analyze the interactions between the fundamental bosons. One result is that as the critical surface is approached small amplitude wave with frequency $\omega$ can decay into 3 fundamental bosons with lifetime [5]:

$$\frac{\hbar}{\tau} = \frac{1}{3\pi^2}\left(\frac{m}{\hbar^2}\right)^3\left(\frac{P_c}{\rho_c}\right)^2 (\hbar\omega)^2 \quad . \qquad (5)$$

When $\tau \ll \tau_0$ an incoming wave will decay with almost 100% probability. Evidently the critical surface is "black" for high frequency modes, but semi-transparent for low frequency modes. At sufficiently high temperatures the critical surface is opaque at all frequencies. If the temperature is not too high the frequency dependent transmission properites of a closed critical surface could serve as an experimental signature for the existence of such a surface [7]. Another signature would be the large heat capacity associated with the critical surface.

It is straightforward to evaluate the thermal energy using the dispersion relation (3) together with $c_s = z/\tau_0$. The result that the thermal energy stored inside the critical surface is [5]

$$U = 7.6 \times 10^{-4} mc^2 \left(\frac{T}{T_H}\right)^3, \qquad (6)$$

where

$$T_H = \frac{\hbar c}{4\pi k_B R_c}. \qquad (7)$$

$T_H$ is a characteristic temperature below which excitations inside the critical surface freeze out due to quatum effects. Numerically $T_H \approx 16(1\text{fermi}/R_c)$ MeV. The internal energy (6) differs from that stored in in a blackbody hohlraum with radius $R_c$ by a factor approximately equal to $mc^2/k_B T$. As shown in ref. [5] this thermal energy is not uniformly distributed throught the volume inside the critical surface as it would be for a hohlraum, but instead is concentrated near to the radius $R_c$. The entropy associated with thermal energy is $U$ is:

$$S = .001 k_B mc^2 \left(\frac{4\pi R_c}{\hbar c}\right)\left(\frac{T}{T_H}\right)^2. \qquad (8)$$

If we compare this entropy with the Bekenstein bound $S_{max}/k_B = 4\pi R_c U/\hbar c$ [8] we find $S = 1.5 \times (T_H/T) S_{max}$. The temperatures reached in RHIC collisions are thought to be close to the thermodynamic critical temperature $T_c$, which is estimated to be about 175 MeV [2]. Therefore the entropy inside the crtical surface will approach the Bekenstein bound if the freeze-out temperature $T_H$ approaches $T_c$. If we identify $R_c$ as the radius of an event horizon, then $T_H$ becomes the Hawking temperature [9].

Measurements of the pion correlation length [10] for QGPs created in RHIC collisions suggest that the size of the QGP at the time of maximum density typically lies in the range 1-2 fermis. However, it may be possible to choose parameters for the colliding heavy ions so that the size of a critical surface created in a relativistic collision is much smaller than the size of the QGP, so that $T_H$ will comparable to $T_c$. Of course, because the thermodynamic QGP critical temperature doesn't depend on the size of the critical surface, a QGP critical surface may be able to faithfully represent the behavior of quantum gravity near an event horizon for only a very narrow choice of intial conditions. On the other hand, we will now argue that the very existence of a QGP critical point has a deep connection with quantum gravity.

In a Bose fluid with a critical point the fundamental bosons reveal themselves both via the decay of collective excitations and the numerical value of the critical surface specific heat and entropy. Of course, this begs the question as to why the fundamental degrees of freedom of a QGP should be regarded as massive bosons  A more sophisticated view is to regard the dominent fundamental degrees of freedom in a hot QGP as color magnetic or dyonic solitons connected by strings [11]. This is a point of view that seems to be roughly consistant with the data on RHIC collisions. Inside a QGP the strings connecting the magnetic solitons will have a typical length depending on the temperature, and near to $T_c$ the magnetic solitons can be simply be regarded as non-relativistic particles with some characteristic mass $m$. In any case the behavior of the speed of sound near to the critical surface in our bosonic model (Fig. 2) is very similar to what is predicted by lattice gauge calculations [12]. We will assume in the following that the crtical point of the bosonic model for a QGP is a stand-in for what in reality is a continuous transition between a QGP state with a high density of magnetic solitons connected with strings and a QGP state with a low density of paired magnetic solitons (cf. Fig 1).

The expected appearance of magnetic solitons connected by strings near to the critical point of a QGP is reminiscent of D-brane representations for a black membrane in 5-dimensional anti-de Sitter (AdS$_5$) space [13], and suggests that a closed critical surface in a QGP might indeed serve as a model for what an event horizon in 4-dimensions should look like in string theory. It may be worth noting in this connection that the existence of a quantum critical point in the phase diagram of a QGP can be inferred from AdS/CFT duality provided the AdS$_5$ space is equipped with a suitable non-conformal potential for the dilaton field and contains a black membrane [14]. At the critical point itself the black membrane in the dual AdS$_5$ space is replaced by an assembly of string theory D-branes. Of course, identifying the thermodynamic critical point with a breakdown in a classical description of AdS$_5$ space outside a black membrane doesn't neccesarily imply that the structure of a closed critical surface in a QGP is a representation for the quantum structure of an event horizon in ordinary 4-dimensional space-time. On the other hand, we can offer an alternative explanation why this is so based on the fact that some forms of gravity in 4-dimensional space-time can also be described as a kind of quantum fluid. In particular, building on the old idea [15] that curved twister spaces should play a fundamental role in a quantum theory of gravity, it has been shown [16-17]. that one can construct a quantum model for certain kinds of 4-dimensional manifolds where the fundamental degrees of freedom are SU(∞) magnetic solitons.

In these twister inspired models for space-time one imagines that 4-dimensional space-time can be represented as a foliation of 2-dimensional surfaces, and that on each surface there is a quantum fluid of nonabelian anyons described by the nonlinear Schrodinger equation of the form:

$$i\hbar \frac{\partial \Phi}{\partial t} = -\frac{1}{2\bar{m}} D^2 \Phi + e[A_0, \Phi] - g[[\Phi^*, \Phi], \Phi] \ , \tag{9}$$

where $D = \nabla - i(e/\hbar c)[A,$ and $A_0$ and $A$ is the gauge potential for an SU(N) Chern-Simons gauge field. In this theory the wavefunction $\Phi$ and potentials $A_0$ and $A$ are N x N SU(N) matrices. As shown in ref. {16} in the limit N→ ∞ Eq. (9) becomes the "Heavenly Equation" for a self-dual Einstein space. It can be shown [17-18] that there are also "ambi-twister" solutions to Eq. (9) where the phase of the wavefunction is expressed as a superposition of the phases for self-dual or anti-self-dual SU(∞) magnetic solitons, which were christened "chirons" in ref. [16]. The effective action corresponding to a superposition of the self-dual and anti-self-dual phases is

$$W = \sum_{<ij>} \pm \tanh^{-1}\left(\frac{u_i - u_j}{R_{ij}}\right), \quad (10)$$

where the sum runs over all pairs of chirons, $R_{jk}^2 = (u_i - u_j)^2 + 4(z_j - z_k)(\bar{z}_j - \bar{z}_k)$, $u_k$ is a coordinate specifying the layer containing the chiron, $z_j$ is the position of a chiron within the layer, and the ± sign specifies whether the magnetic charge of the two chirons is the same or opposite. This effective action resembles the condensate phase of a configuration of 2-dimensional XY vortices, which suggests that the properties of a chiron fluid may be related to the properties of a 2-dimensional Coulomb gas. Indeed, as shown in ref. [18] substituting the phase (10) into the Polyakov action for a 2-dimensional scalar conformal field theory yields a partition function resembling that for a discrete 2-dimensional Coulomb gas, with the spacings $u_i - u_j$ between layers playing the role of the lattice spacings. This similarity shows that a chiron fluid containing a mixture of self-dual and anti-self-dual solitons will exhibit a 3-dimensioal phase transition resembling the Kosterlitz –Thouless (KT) condensation of vortex and anti-vortex pairs in the 2-dimensional XY model. In the case of the chiron fluid the low temperatlure side of this phase transition represents a classical 4-dimensional manifold which can be constucted using the ambi-twister method [19].

In a 2-dimensional Bose superfluid the KT transition would take place at the temperature [20]:

$$k_B T_{KT} = \pi \hbar^2 \rho_s / 2m^* \tag{11}$$

where $\rho_s$ is the 2-dimensional density of bosons at the critical point and m* is the mass of the boson. If we assume that the number of particles in the critical layer will be on the order of $S_c/k_B$. If we assume that the critical surface is spherical with radius $R_c$, and use the the entropy given in Eq. (8) to estimate to estimate $S_c$, then we arrive at the result that $T_{KT} \approx 3 T_H$. Even though it is only a rough approximation to regard the transition in a chiron fluid with effective action (10) as a 2-dimensional KT transition, it is noteworthy that we obtain an estimate for the critical temperature is not far from the Hawking temperature. Thus a chiron fluid has the remarkable property that the critical temperature seems to vary with the size of the critical surface in exactly the same way that the Hawking temperature depends on the radius of an event horizon. Of course this may mean that there is a profound difference between a QGP and a chiron fluid in that in the QGP case the critical temperature doesn't depend on the size of the QGP or critical surface in the thermodynamic limit. It should be noted though that neither lattice gauge calculations nor AdS/CFT duality can predict the actual value of the critical temperature in a dynamical QGP, so it may turn out that for the finite size QGPs created in RHIC collisions the critical temperature does depend on the size of the critical surface after all. In any case it is remarkable that the critical temperature of QGPs created in relativistic heavy ion collisions will apparently be not far from the KT critical point (11).

Perhaps the most enigmatic aspect of our attempt to identify a closed critical surface in either a QGP or a chiron quantum fluid with a space-time event horizon in 4-dimensions concerns the behavior of the fluid inside the critical surface. Naively one might expect that the fluid inside the critical surface should correspond to the interior space-time for a black hole. However, such an interpretation for the interior fluid clearly conflicts with what is predicted by the bosonic model defined in Eq. (1).

If we ignore the 2nd term in Eq. (2) then this equation resembles the classical equation for wave propagation in a space-time with metric

$$ds^2 = dx^2 + dy^2 + dz^2 - c_s^2 dt^2 \ . \tag{12}$$

On the outside of the critical surface the metric (12) has the same form as the Schwarzschild metric just outside its horizon if we make the identification $c_s = c^3z/4GM$, where $M$ is the black hole mass. On the inside this metric resembles the interior de Sitter metric near its horizon if we make the identification $c_s = (\Lambda/3)^{1/2}z$, where $\Lambda$ is the cosmological constant. In the context of classical general relativity such a configuration would be stable only if the pressure inside the horizon is negative [21], which is presumably not the case for a quark-gluon plasma. However, it may be worth noting that the pressure of the QGP inside the critical surface is significantly lower than what would be expected for a an ideal gas of quarks and gluons. In the quark bag model for a gas of quarks [22] this decrease in the pressure is regarded as being due to the contribution of a vacuum energy with negative pressure. Thus the departure in the behavior of the QGP inside the critical surface from what is expected in general relativity may just reflect the suggestions in refs. [5, 23] that the failure of general relativity at the event horizon itself changes the space-time over the entire volume inside the critical surface. In fact these suggestions are in accord with both an analysis of the behavior of quantum gravity near to an event horizon in 4-dimensions based on the conformal anomaly [24], as well as an analysis of Green's functions in infinite de Sitter space [25].

### Acknowledgements

The author is grateful to A. Kerman, P. Mazur, E. Mottola, N. Snyderman, R. Soltz, and J. Zaanen for helpful discussions.